\documentstyle{elsart}
\input epsf 
\begin{document}
\newcommand{\lesssim}{\,\vcenter{\hbox{$\buildrel\textstyle<\over\sim$}}}
\newcommand{\gtrsim}{\,\vcenter{\hbox{$\buildrel\textstyle>\over\sim$}}}

\begin{frontmatter}
\title{Near-Surface Long-Range Order at the Ordinary Transition:
Scaling Analysis and Monte Carlo Results}
\author{Peter Czerner and Uwe Ritschel}
\address{Fachbereich Physik, Universit\"at GH Essen, 45117 Essen
(F\ R\ Germany)}
\begin{abstract}
Motivated by recent experimental activities on surface critical phenomena,
we present a detailed theoretical study of the near-surface behavior of the
local order parameter $m(z)$ in Ising-like spin systems.
Special attention is
paid to the {\it crossover} regime 
between ``ordinary'' and ``normal'' transition in the three-dimensional
semi-infinite Ising model,
where a finite
magnetic field $H_1$ is imposed on the surface which itself
exhibits a reduced tendency to order spontaneously.
As the theoretical foundation,
the spatial behavior of $m(z)$ is discussed by means of 
phenomenological scaling arguments, and a finite-size scaling analysis
is performed. Then we present
Monte Carlo results for $m(z)$
obtained with the Swendsen-Wang algorithm.  
In particular the sharp power-law
increase of the magnetization, $m(z)\sim H_1\,z^{1-\eta^{ord}_{\perp}}$,
predicted for a {\it small} $H_1$ by previous work of the authors, is
corroborated by the numerical results.
The relevance of these findings for experiments
on critical adsorption in systems where a small effective
surface field occurs is pointed out.
\end{abstract}
\end{frontmatter}
\section{Introduction}\label{intro}
A great deal of current experimental activity concentrates on the
investigation of critical phenomena near surfaces. After
the impressive confirmation of the theoretical 
predictions  \cite{binder,diehl,didi,diwa}
in experiments with binary alloys \cite{mail},
the more recent experimental efforts
focus on binary mixtures near their consolute
point \cite{law,beysens,franck} and near-critical
fluids \cite{findenegg}.
In most of these experiments the
order parameter, the concentration difference in fluid
mixtures or the density difference between liquid and gaseous
phase in fluids,
plays a central role. For instance the reflectivity and
ellipticity
measured in light-scattering experiments are directly
related to the order-parameter profile \cite{reflec,ellipt}.
Hence, quantitative  information about the local order parameter near
the boundary
is necessary for the interpretation of the experimental data.

While a very well-developed theory exists for the
individual {\it surface universality classes}, corresponding to
fixed point of the
renormalization-group flow, the picture in the crossover
regions between the fixed points is less complete. The
experiments are generically not carried at the fixed points,
however, and so a detailed understanding of the
crossover region is particularly important.  

Consider for example the semi-infinite three-dimensional (3-$d$)
Ising system with spin-spin
interaction $J$.  
In this model the influence of the surface
is usually taken into account by means of 
a modified exchange interaction $J_1$ in the surface
and a magnetic field $H_1$ imposed on the
surface spins  \cite{binder,diehl}. While the former models
modifications due to the surface within the critical medium, the
latter represents the influence of the adjacent (noncritical)
medium, as the container wall for example, on the system.

For a brief (and necessarily incomplete)
summary on surface critical phenomena,
let us first set $H_1=0$. Then at the bulk critical point $T_c$
the tendency to order near the surface
can be reduced, increased,
or unchanged compared with the bulk. Which case is realized depends on the ratio $J_1/J$.
At a particular value, $J_1^{sp}\simeq 1.5\,J$ \cite{binder,labi,ruge} 
the third case is realized. 
This corresponds to the {\it surface universality class} of the
``special transition''.
For $J<J_1^{sp}$ the surface
has a reduced tendency to order but nevertheless
becomes (passively) ordered at the bulk phase transition. 
In the opposite case, $J>J_1^{sp}$, the surface orders at a temperature
{\it above} $T_c$, and {\it at} $T_c$ the bulk undergoes
a phase transition in the presence of an already ordered surface.
From the viewpoint of the renormalization group 
the special transition is an unstable fixed point \cite{diehl}.
For a start value $J_1 < J_1^{sp}$ the (running)
surface coupling is driven to the stable
fixed point $J_1=0$ corresponding to the universality class
of the ``ordinary transition". For $J_1 > J_1^{sp}$ it is driven to
$J_1=\infty$, again a stable fixed point,
corresponding to the universality class of the 
``extraordinary transition''  \cite{diehl,smock}.

Next we consider the effects of $H_1$
in a system with $J_1<J_1^{sp}$. This is, for example, 
the situation generically met in experiments with
binary fluids. In particular we are interested in the behavior of the
order parameter in this situation.
The universality classes are determined by the fixed-point
values $H_1=0$ and
$H_1=\infty$ of the renormalization-group transformations.
For $H_1=0$, at the ordinary transition,
the order parameter $m(z)$
simply vanishes since, in terms of Ising spins, the
symmetry under the reversal $s_i\to -s_i$ is not 
broken, neither
in the bulk nor in the surface. For $H_1=\infty$ 
the universality class is called the ``normal transition''.
The normal transition is known to be equivalent to
the extraordinary transition \cite{bray,dibu}. In both cases $m(z)$
starts from a large
$m_1$ at the surface and then decays to the
bulk equilibrium value (being zero for $T\ge T_c$ and nonzero
for $T<T_c$). At $T_c$, i.e. for infinite correlation length $\xi$,
the decay is described by a universal
power law $m\sim z^{-\beta/\nu}$ for macroscopic
distances $z$. For instance for the 3-$d$ Ising model
$\beta/\nu \simeq 0.52$  \cite{ferlan}.
For $T\neq T_c$ a crossover to the
exponential decay $\sim \exp(-z/\xi)$ takes place in a distance
$z\simeq \xi$ from the surface.

What happens in the crossover region between $H_1=0$ and
$H_1=\infty$?
Mean-field theory predicts a profile that starts
from some finite $m_1$ and then monoto\-nously decays to
the equilibrium value.
In Ref.\,\cite{czeri}
the present authors have shown that,
contrary to the naive (mean-field)
expectation, fluctuations may cause the order parameter to steeply
{\it increase} to values $m(z)\gg m_1$
in a surface-near regime. This growth
is described by a universal power law
\begin{equation}\label{power}
m(z) \sim H_1\,z^{\kappa}\quad \mbox{with} \quad \kappa=1-n_{\perp}^{ord}\>,
\end{equation}
where $\eta^{ord}_{\perp}$ is the anomalous dimension
pertaining to the ordinary transition (governing the decay of
correlations in the direction perpendicular to the surface \cite{diehl}).
For instance for the 3-$d$ Ising model $\kappa\simeq 0.21$.

The scenario for the crossover between ordinary and normal
transition developed in \cite{czeri} 
is the following: At bulk
criticality and $J_1<J_1^{sp}$ for any finite $H_1$ the order-parameter
profile
increases up to a certain length scale $l^{ord}$ and then crosses over
to the power law $\sim z^{-\beta/\nu}$. The scale $l^{ord}$ 
is given by an {\it inverse} power of $H_1$ and, thus, becomes smaller
for increasing $H_1$ such that in the limit $H_1\to\infty$
the maximum has moved to the surface. In this limit only
the previously mentioned monotonous power-law decay characteristic
for the normal transition is left over. A qualitative
sketch of typical crossover profiles is shown in Fig.\,1.
In this plot the axes are logarithmic and
both $m(z)$ and $z$ are measured in arbitrary units. The individual
curves have the correct asymptotics, $m(z)\sim z^{0.21}$ for
$z\to 0$ and $m(z)\sim z^{-0.52}$ for $z\to \infty$. However,
the (yet unknown) real crossover function is replaced by a simple
substitute.\\[2mm]  
\begin{figure}[b]
\def\epsfsize#1#2{0.6#1}
\hspace*{2cm}\epsfbox{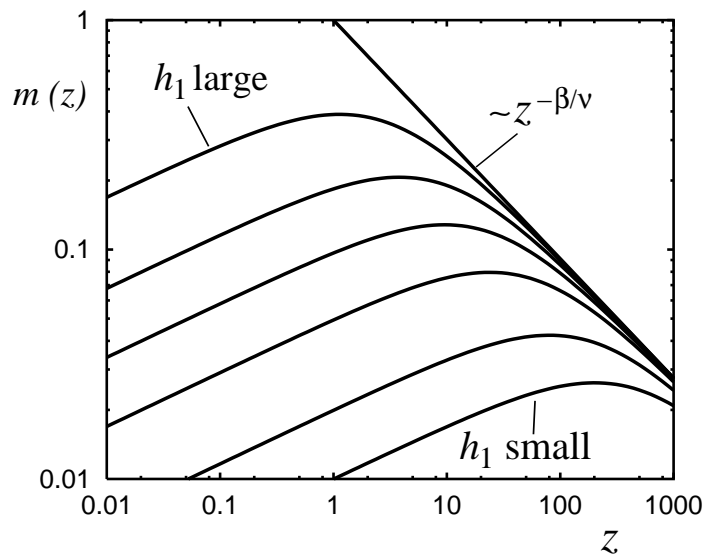}\\[-2cm]
\caption{Qualitative sketch of order-parameter profiles in the
crossover regime between ordinary and normal transtion in
double-logarithmic representation. Both $z$ and $m$ are given in
arbitrary units.}
\end{figure}

In Ref.\,\cite{twod} it was demonstrated by means of MC simulations
and the comparison with exact results
that also in the 2-$d$
Ising model the crossover between ordinary and normal transition
is qualitatively of the same form as in $d=3$. However,
the simple power law (\ref{power}) is modified by a logarithm in $d=2$.
The main purpose of the present work is to verify the
results for the order parameter obtained in \cite{czeri},
where scaling and heuristic arguments were used and a quantitative
calculation in the framework of renormalization-group improved
perturbation theory was performed,
by means of Monte Carlo (MC) simulation explicitly for the
three-dimensional system.

The MC studies devoted to (static) critical phenomena near
surfaces that are contained
in the literature
concentrated mainly on the dependence of thermodynamic variables
on $J_1/J$ and, in particular, on critical adsorption
for large $J_1$  \cite{extraord} as well as on the precise location of 
$J_1^{sp}$ \cite{ruge,bila1}. $H_1$ was set to zero
in these studies. 
Numerical studies of the influence of $H_1$ concentrated
on the surface layer magnetization  \cite{bila2} and,
in the context of wetting, on systems {\it below}
the critical temperature 
 \cite{wett}.
To our knowledge, there is no work in the literature
where order-parameter profiles
at or above $T_c$ with non-vanishing $H_1$ were studied by MC methods
and which could have been directly compared with the
analytic results reported in Ref.\,\cite{czeri}.

The rest of this paper is organized as follows: In Sec.\,\ref{two} we
summarize and supplement the main results of \cite{czeri},
especially the phenomenological scaling analysis which allows
to make quite precise predictions for $m(z)$ in the crossover
regime. In Sec.\,\ref{three}
our MC procedure, essentially the Swendsen-Wang algorithm slightly
modified to allow for the inclusion of a surface field, is described. 
The MC results are presented in Sec.\,\ref{results}.
Eventually, the last section contains besides a short summary
remarks on the
relevance of our results for experiments.

\section{Theory}\label{two}
\subsection{Model}\label{model}  
We consider the semi-infinite Ising system with a free boundary
on a plane square lattice.
The exchange coupling between direct neighbors in the bulk
is $J$. In the surface the nearest-neighbor coupling is $J_1$.
A surface magnetic field $H_1$ is
imposed on the boundary spins and bulk magnetic fields are set to
zero such that the Hamiltonian of the model reads
\begin{equation}\label{ising}
{\cal H}_{\rm Ising}= -J\!\sum_{<ij>\in V}\,s_is_j-J_1\!\sum_{<ij>\in\partial V}\,s_is_j-H_1\sum_{i\in\partial V}\,s_i\>,
\end{equation}
where
$\partial V$ and $V$ stand for the boundary and the rest of the
system (without the boundary), respectively.
Below we mainly work with the dimensionless variables
\begin{equation}\label{Kandh1}
K=J/k_BT\>,\quad K_1=J_1/k_BT\>,\quad\mbox{and}\quad  h_1=H_1/k_BT\>.
\end{equation}
For the (reduced) critical bulk coupling 
we took $K_c\equiv J/k_BT_c=0.22165$ from the literature \cite{ferlan}.
The value of $K_1$ that corresponds to the special transition
is $K_1^{sp}=1.5\,K_c$  \cite{bila1,ruge}. 

\subsection{Scaling Analysis}\label{scalan}

In the critical regime
thermodynamic quantities are described
by homogeneous functions of the scaling fields.
As a consequence,
the behavior of the local magnetization under
rescaling of distances should be described by
\begin{equation}\label{scal}
m(z,\tau,{h}_1)\sim b^{-x_{\phi}}\,m(zb^{-1},\,\tau b^{1/\nu},\,
{h}_1\,b^{y^{ord}_1})\>,
\end{equation}
where $x_{\phi}=\beta/\nu$ and
$y^{ord}_1=\Delta^{ord}_1/\nu$ are the scaling dimensions
of the equilibrium magnetization $m(z\to \infty)$
and the surface field $h_1$, respectively \cite{diehl}.
In general the surface exponents have different values
for different surface universality scales \cite{diehl}, and
so these quantities are additionally marked by `{\it ord} ' for
belonging to the ordinary transition. 
The (MC) literature values for the 3-$d$ Ising model are
$x_{\phi}=0.518(7)$  \cite{ferlan} and $y_1^{ord}=0.73$.
The value $y_1^{ord}$ was obtained by
employing the scaling relation $y_1+x_1=d-1$ together with
the recent Monte Carlo result $x_1^{ord}=\beta_1^{ord}/\nu=1.27$ \cite{ruge2}.

Removing the arbitrary rescaling parameter $b$ in
Eq.\,(\ref{scal}) by setting it
$\sim z$, one obtains the scaling form of the magnetization
\begin{equation}\label{scalm}
m(z,\tau,h_1)\sim z^{-x_{\phi}}\,{\cal M}(z/\xi, z/l^{ord})\>,
\end{equation}
where
\begin{equation}\label{length}
l^{ord}\sim h_1^{-1/y_1^{ord}}
\end{equation}
is the length scale determined by the surface field. The second length
scale pertinent to the semi-infinite system and occurring in
(\ref{scalm}) is the bulk correlation length $\xi=\tau^{-\nu}$.
Regarding the interpretation of MC data, which are normally obtained
from finite lattices, one has to take into account a third
length scale, the characteristic dimension $L$ of the system,
and a finite-size scaling analysis
has to be performed. The latter will be described
in Sec.\,\ref{fisi}.

Going back to the semi-infinite case and setting
$\tau=0$,
the only remaining length scale is $l^{ord}$, and
the order-parameter profile can be written in the
critical-point scaling form
\begin{equation}\label{h1}
m(z,{h}_1)\sim z^{-x_{\phi}}\,{\cal M}_c(z/l^{ord})\>.
\end{equation}
As said above, for $z\to \infty$ the magnetization decays as $\sim
z^{-x_{\phi}}$
and, thus, ${\cal M}_c(\zeta)$ should approach a constant for
$\zeta\to \infty$.
In order to work out the {\it short-distance}
behavior of the scaling function ${\cal M}_c(\zeta)$,
we demand that $m(z)\sim m_1$ as $z\to 0$.
This means that in general, in terms of macroscopic quantities,
the boundary value of $m(z)$ is {\it not} $m_1$. If
the $z$-dependence of $m(z)$ is described
by a power law, it cannot approach any value
different from zero or infinity as $z$ goes to zero.
However, the somewhat weaker relation symbolized by ``$\sim$'' should hold,
stating that the respective
quantity asymptotically (up to constants) ``behaves as" or
``is proportional to".
This is in accord with and actually motivated by the
field-theoretic short-distance expansion  \cite{syma,diehl}, where
operators near a boundary are represented in terms of
boundary operators multiplied by $c$-number functions.

In the case of the 3-$d$ Ising model the foregoing
discussion leads to the conclusion
that $m(z)\sim h_1$ because the ``ordinary''
surface with $K_1<K_1^{sp}$ is paramagnetic
and responds linearly to a small magnetic field \cite{bray}.
This is different in $d=2$ where an additional
logarithmic factor occurs  \cite{fishau}, $m_1\sim h_1\,\mbox{ln}\,h_1$,
and this logarithm also leaves its fingerprint on the near-surface
behavior of the magnetization  \cite{twod}.

The immediate consequence of the simple
{\it linear} response in $d=3$ for the scaling function ${\cal M}_c(\zeta)$ occuring in (\ref{h1})
is that it has to behave as $\sim \zeta^{y^{ord}_1}$ in the small-$\zeta$ limit. After inserting
this in (\ref{h1}),
we obtain that the exponent governing the short-distance
behavior of $m(z)$ is given by the difference between
$y_1^{ord}$ and $x_{\phi}$, such that for $z\ll l^{ord}$ the
magnetization is described by
\begin{equation}\label{power2}
m(z)\sim h_1 \,z^{y^{ord}_1-x_{\phi}}\>.
\end{equation}
Using the
scaling relations $\eta_{\perp}=(\eta+\eta_{\parallel})/2$
and $y_1=(d-\eta_{\parallel})/2$ \cite{diehl}, 
we eventually obtain $\kappa$ as expressed in (\ref{power}).

In the mean-field approximation
the result for $\kappa$ is zero. Thus, in this case
one really has $m(z\to 0)=m_1$ and the monotonously decaying order-parameter profile
mentioned earlier.
However, a positive value is obtained when fluctuations are taken into account
below the upper critical dimensionality $d^*=4$.
Taking $y_1^{ord}\simeq 0.73$ from above, the
value for $\kappa$ is 0.21.

Phenomena to some extent analogous to the ones discussed above were
reported for the crossover between {\it special} and
normal transition \cite{brezin}.
Also near the special transition the
surface field $h_1$ gives rise to a length scale. However, the
respective exponent, the analogy to $\kappa$,
is negative, and, thus, one
encounters a profile that {\it monotonously} decays for all
(macroscopic) $z$, with different power laws
in the short-distance and the long-distance regime and
a crossover at distances comparable to the length scale
set by $h_1$. However,
{\it non-monotonous} behavior in the crossover
region as described above for $m(z)$ is a common feature in the case
of the energy density in $d=2$  \cite{mifi} as well as in
higher dimensionality \cite{eisenriegler}.

\subsection{Relation to Critical Dynamics}\label{critdyn}
The {\it spatial} variation of the
magnetization discussed so far
strongly resembles the {\it time} dependence of the
order parameter in relaxational processes at the critical point.
If a system with nonconserved order parameter (model A) is quenched from a
high-temperature initial state to the critical point, with a small
initial magnetization $m^{(i)}$, the order parameter behaves as $m \sim
m^{(i)}\,t^{\theta}$  \cite{jans}, where the short-time exponent $\theta$
is governed
by the difference between the scaling dimensions of initial
and equilibrium magnetization divided by the
dynamic (equilibrium) exponent \cite{own1}. Like the exponent $\kappa$
in (\ref{power}), the exponent
$\theta$ vanishes in MF theory, but becomes positive
below $d^*$. For example, its value in the 3-$d$ Ising model with Glauber
dynamics is $\theta=0.108$  \cite{theta}.

The high-temperature initial state of the relaxational process is to
some extent analogous
to the surface that strongly disfavors the order and that (for
$h_1=0$) belongs to the universality class of the ordinary transition.  
Further expanding this analogy, heating a system
from a low-temperature (ordered) initial state to
the critical point would be similar to the situation
at the extraordinary transition. Eventually, analogous to
the special transition would be a ``relaxation process'' that starts
from an equilibrium state at $T_c$. 

\subsection{Heuristic Argument}\label{heuristic}
There is also a heuristic argument for the growth of the magnetization
in the near-surface regime expressed in (\ref{power}).
As said above, a small $h_1$ generates a surface magnetization $m_1\sim h_1$.
Regions that are close to the surface 
will respond to this surface magnetization by ordering as well.
How strong this
influence is depends on two factors.

First, it is proportional to
the correlated area in a plane parallel to the surface at a distance $z$.
While correlations in the surface are asymptotically (for $J_1\to 0$) suppressed, 
for $z> 0$ the range of correlations between spins located in a plane parallel to
the surface in a distance $\rho$ from each other can be regarded as finite, because
for $\rho> z$ the parallel correlation function is governed by surface
exponents and decays much faster than in the bulk. Hence the corresponding
effective correlation length, $\xi_{\parallel}(z)$, should behave as $z$.
Referring once more to
critical dynamics as discussed in the previous section, $\xi_{\parallel}(z)$
is analogous to the time dependent (growing)
correlation length $\xi(t)\sim t^{1/\zeta}$
(where $\zeta$ here stands for the dynamic equilibrium exponent).
  
Second, it
depends on the probability that a given spin orientation has
``survived'' in a distance $z$. For small $h_1$ and $z<l^{ord}$,
the latter is governed by the {\it perpendicular}
correlation function $C(z)\sim z^{-(d-2+\eta_{\parallel}^{ord})}$.
Taking into account both factors, we obtain
\begin{equation}\label{heuris}
m(z)\sim h_1\,C(z)\,\xi_{\parallel}^{d-1}=h_1 z^{1-\eta_{\perp}^{ord}},
\end{equation}
the short-distance power law reported in (\ref{power}).
Qualitatively speaking, 
the surface when carrying a small $m_1$ induces a much larger
magnetization in the adjacent layers, which are much more
susceptible and capable of responding with a
magnetization $m\gg m_1$. 

This simple picture for the anomalous short-distance behavior holds
for dimensions $2<d<4$. At and above the the upper critical dimension $d^*=4$, where
the mean-field theory starts to provide the correct description, 
the power-law growth of magnetization is not observed, 
since there the increase of the correlated surface
area is  compensated by the decay of the
perpendicular correlation function. In the case of the two-dimensional
Ising model the assumption that $m_1\sim h_1$ is no longer valid,
and logarithmic terms occur \cite{twod}.
    
\subsection{Modifications at $T\neq T_c$}\label{temp}
The phenomenological scaling analysis 
presented above can be straightforwardly extended
to the case $\tau>0$. In $d>2$, we may assume that the
behavior near the surface for $z<<\xi$ is unchanged compared
to (\ref{power}), and, thus,
the increasing profiles are also expected slightly above the
critical temperature. The behavior farther away from the surface depends
on the ratio $l^{ord}/\xi$.
In the case of $l^{ord}>\xi$ a crossover
to an exponential decay will take place for $z\simeq \xi$
and the regime of nonlinear decay does not occur.
For $l^{ord}< \xi$ a crossover
to the power-law decay $\sim z^{-\beta/\nu}$ takes place
and finally at $z\simeq \xi$ the exponential behavior sets in.

An interesting phenomenon can be observed in the case
$\xi <l^{ord}$. As discussed above,
$m(z)$ then never reaches the regime with power-law decay, but
crosses over from the near-surface increase directly to the exponential
decay. Since the region where $m(z)$ grows extends up to the
distance $\xi$, the magnetization in the maximum has roughly the
value $m_{max} \simeq \xi^{\kappa}$. Now,
the amplitude of the exponential decay should behave
as $\sim m_{max}$ such that
for $z\gg \xi$ we have
\begin{equation}\label{expo}
m(z)\sim h_1\,\xi^{\kappa}\,\exp(-z/\xi)\>.
\end{equation}
In other words, in the case $\xi < l^{ord}$ the short-distance exponent
$\kappa$ not only governs the behavior of $m(z)$ near the surface,
but also leaves its fingerprint much farther away from the surface
in form of an universal dependence of the {\it amplitude} of the
exponential decay on the correlation length $\sim \xi^{\kappa}$.
Nothing comparable occurs when $\xi =\infty$
(compare Sec.\,\ref{scalan} above). When $l^{ord}$ is the only
scale, all profiles approach the same
curve $m(z)\approx {\cal A}\, z^{-\beta/\nu}$
for $z\gtrsim l^{ord}$, with an amplitude ${\cal A}$ {\it independent}
of $h_1$. An analogous phenomenon, termed ``long-time memory'' of
the initial condition, does also occur in critical
dynamics for $T\ge T_c$  \cite{own2}.

Below the critical temperature (and near the ordinary transition), the short-distance
behavior of the order parameter is also described by a power
law, this time governed by a different exponent, however  \cite{gompper}.
The essential
point is that below $T_c$ the surface orders spontaneously even for $h_1=0$.
Hence, in the scaling analysis the scaling dimension of $h_1$
has to be replaced by the scaling dimension of $m_1$, the
conjugate density to $h_1$,
given by $x_1^{ord}=\beta^{ord}_1/\nu$ \cite{diehl}. The exponent that describes
the increase of the profile is thus
$x_1^{ord}-x_{\phi}$  \cite{gompper}, a number that even in
mean-field theory is different from zero ($=1$) and for the $3-d$ Ising model
its value is 0.75.

\subsection{Finite Size Scaling}\label{fisi}
In order to assess the finite size effects to be
expected in the MC simulations, we have to take into account the
finite-size length scale $L$. The latter is proportional to the linear
extension of the lattice (called $N$ below).
The generalization of (\ref{scal}) reads \cite{fisi}
\begin{equation}\label{scalfs}
m(z,\tau,{h}_1,L)\sim b^{-x_{\phi}}\,m(zb^{-1},\,\tau b^{1/\nu},\,
{h}_1\,b^{y_1^{ord}}, Lb^{-1})\>,
\end{equation}
and proceeding as before, we obtain as the generalization
of (\ref{scalm}) to a system of finite size:
\begin{equation}\label{scalmfs}
m(z,\tau,h_1,L)\sim z^{-x_{\phi}}\,{\cal M}(z/\xi, z/l^{ord},z/L)\>.
\end{equation}
Thus even at $T_c$ there are two macroscopic length scales,
on the one hand $L$ (imposed by
the geometry)
and on the other hand $l^{ord}$ (the scale set by $h_1$).

It is well known that for large $z\gtrsim L$ we have to expect
an exponential decay of $m(z)$ on the
scale $L$. In the opposite limit,
when $z$ is smaller than both $L$ and $l^{ord}$, we expect
the short-distance
behavior (\ref{power}) to occur. That this expectation turns out to be correct
is the necessary condition for observing (\ref{power}) in MC simulations.
Stated in terms of correlation lengths it means that as long as $\xi_{\parallel}$
(cf. discussion in Sec.\,\ref{heuristic}) is smaller than $L$ (and the bulk correlation
length $\xi$), the form of the profile is unchanged compared to the one of
the semi-infinite system (at bulk criticality). In particular it implies that the
surface magnetization $m_1$, whose linear response to $h_1$  was
an important ingredient to our scaling analysis of Sec.\,\ref{scalan}, should not
depend on $L$, as long as $L$ can be regarded as macroscopic. 
   
Farther away from the surface, the form of the profile
depends on the ratio between $l^{ord}$ and $L$.
In the case of $l^{ord}>L$ a crossover
to an exponential decay will take place for $z\simeq L$, and,
analogous to the situation with a finite correlation length, also
in the finite-size system the amplitude of this exponential
decay is governed by the exponent $\kappa$ (compare to
(\ref{expo}) and the discussion in Sec.\,\ref{critdyn}), such that we
have for $z\gg L$
\begin{equation}\label{expol}
m(z) \sim h_1\,L^{\kappa}\,\exp(-z/L)
\end{equation}
Again, analogous finite-size effects were reported also in
relaxation processes near criticality  \cite{own2}. 
In the
opposite case, $l^{ord}<L$, a crossover
to the power-law decay $\sim z^{-x_{\phi}}$ takes place,
followed by the crossover to the exponential
behavior at $z\simeq L$. Thus, qualitatively,
the discussion for systems of finite size is
largely analogous to the one in Sec.\,\ref{critdyn} for
finite $\xi$.

\section{Monte Carlo Simulation}\label{three}
\subsection{System}\label{geom}
The results of the scaling analysis, especially the
short-distance law (\ref{power}), were checked by MC simulations.
To this end we calculated order-parameter profiles for the 3-$d$ Ising model
with uniform bulk exchange coupling $K$ and set
$K_1=0$, corresponding to the fixed-point value of the ordinary
transition.
 
The geometry of the systems studied was that of a rectangular (cuboidal)
lattice with two free surfaces opposite to each other
and the other boundaries periodically coupled.
The surface field $h_1$ was imposed on both free surfaces.
The linear dimension perpendicular to the surfaces
was taken to be two times larger than the lateral extension
in order to keep corrections due to the second surface, the
so-called Fisher-de Gennes effect \cite{fidege},
small \cite{fidege}. Hence, when we talk about a lattice of size
$N$ in this section, we refer to a rectangular system
with $N^2\times 2\,N$ spins.

The distance from the surface is still called $z$ in the following,
although it is clearly an integer quantity,
with $z=0$ corresponding to the location of one of the surfaces.
Order parameter profiles were calculated
by averaging in individual configurations over planes parallel to
the surface and, in turn, we averaged over a large number
of configurations generated by the algorithm described in 
Sec.\,\ref{swewa}. Eventually the symmetry of the system
was used and also the average between the left and right half of
the lattice was taken.

\subsection{Procedure}\label{swewa}

We consider the Ising model, defined by (\ref{ising}).
The aim of the (equilibrium) Monte Carlo procedure is to generate a representative sample of configurations ${\bf s}$
distributed according to the Boltzmann factor $P({\bf s}) \sim \exp \left[{\cal H}({\bf s})/k_BT\right]$  \cite{bihe}.
Further, it must be guaranteed that, starting from any initial configurations, after a reasonable amount of time such a sample can be extracted. The latter is in principle
provided if the algorithm that generates a new configuration ${\bf s}'$ from the old one
satisfies {\it detailed balance}.
In terms of transition probabilities 
$W({\bf s}\!\to\! {\bf s}')$ this condition
can be expressed as
\begin{equation} \label{detbal}
W({\bf s}\!\to \!{\bf s}') \exp\left[-{\cal H}({\bf s})/k_BT\right] = W({\bf s}'\!\to {\bf s}) \exp\left[-{\cal H}({\bf s}')/k_BT\right]\>.
\end{equation}
\par

For practical purposes, however, not any algorithm satisfying (\ref{detbal})
is suitable for MC simulations of critical or near-critical systems.
The reason is that physically meaningful algorithms, like Glauber
and Kawasaki dynamics \cite{bihe},
are greatly hampered by the {\it critical slowing down} upon
approaching the equilibrium.
One way out of this dilemma is the Swendsen-Wang (SW) algorithm \cite{swewa},
which satisfies (\ref{detbal}) but does (probably) not correspond
to a physically meaningful dynamics. 

The SW algorithm generates a transition (or update)
${\bf s}\to {\bf s}'$ between 
spin configurations via connected 
bond clusters. A cluster configuration ${\bf n}$ is constructed 
from ${\bf s}$ by
creating bonds between neighboring spins of {\it equal} sign. Then
these bonds are ``activated'' \cite{janke} with probability 
\begin{equation} \label{bondprob}
p=1-e^{-2K}.
\end{equation}
No bonds are generated between spins of {\it opposite} sign.
As the next step, bond clusters are defined as connected sets
of active bonds.
Also isolated spins are identified as a cluster, such that
eventually each spin belongs to a cluster.

In order to obtain a new spin configuration ${\bf s}'$
from ${\bf n}$, one assigns to all sites of a given cluster
a new spin value with equal probability for each
spin direction (independent of the old spin value).
The probability for the transition ${\bf s}\to {\bf s}'$
\begin{equation} \label{p1}
W({\bf s} | {\bf n} | {\bf s}')=p^b(1-p)^mq^{-N_c}
\end{equation}
where ${\bf n}$ is an intermediate cluster
configuration with $N_c$ clusters, and
$b$ and $m$ are the numbers
of ``active'' and ``inactive'' bonds, respectively. This
transition corresponds to one Monte Carlo sweep.

In order to verify that the algorithm satisfies detailed balance,
we have to consider a transition in the opposite direction.
For the transition ${\bf s}'$ from ${\bf s}$ via the {\it same}
cluster configuration ${\bf n}$, there is a probability 
\begin{equation}
W({\bf s} |{\bf n} | {\bf s}')=p^b(1-p)^{m'}q^{-N_c},
\end{equation}
with the same $b$ and $N_c$ as before. 
However, the number of non-active bonds $m'$ can in general be different, because neighboring clusters can originate from
domains with spins of the same or of different sign,
in both cases leading to the same cluster configuration.  

The total transition probability from ${\bf s}$ to ${\bf s}'$ is given by
\begin{equation}
W({\bf s} ,{\bf s}')=\sum_{{\bf n}}W({\bf s}|{\bf n}|{\bf s}'),
\end{equation}
where the sum runs over all possible intermediate
cluster configurations ${\bf n}$. 
Since the sum $b+m$ is constant for a given spin configuration, it 
is straightforward to show that 
\begin{equation}
\frac{W({\bf s} \rightarrow {\bf s}')}{W({\bf s}' \rightarrow {\bf s})}=(1-p)^{m-m'}.
\end{equation} 
Eventually, taking into account that the energy difference between
${\bf s}$ and ${\bf s}'$ is given by
\begin{equation}
-2J(m-m')=-\Delta {\cal H}\>,
\end{equation}
with (\ref{bondprob}) one obtains the detailed balance relation
(\ref {detbal}).

The algorithm presented so far works as
long as no magnetic fields are imposed on the spins.
To take into account the third term in (\ref{ising})
that describes the influence of
the surface magnetic field $H_1$, we follow Wang \cite{wang}
and introduce a layer of
``ghost'' spins next to the surface
that couple to the surface spins only.
The ``ghost'' spins all point in the direction of $H_1$
 and couple to the ``real'' spins with coupling strength equal to $H_1$.
If one or more ``active'' bond between
a surface and a ghost spin exist, the
cluster has to keep its old spin
when the system is updated. This prescription preserves
detailed balance.
In the practical calculation this
rule was realized by a modified (reduced) spin-flip probability
\begin{equation}
p(n_s)=1-\frac{1}{2} \,\exp(-2\,h_1\, n_s)
\end{equation}
for clusters pointing in the direction of $h_1$, where $n_s$ is the number of {\it surface}
spins contained in the cluster. For clusters pointing in opposite direction the probability
has to remain unchanged (equal to 1/2).

In order to obtain an equilibrium sample, we
discarded several hundred -- the precise number depended on the
size of the system -- configurations after the start of the run.
To keep memory consumption low, we used multispin-coding techniques, i.e.
groups of 64 spins were coded in one long integer. All calculations
were run on
a Silicon Graphics computer (Power Challenge)
with four Risk 8000 processors.
To obtain a profile with reasonable statistics for our largest system
(N=256) took about one week of (single-processor) CPU time.

\subsection{Results}\label{results}
From the magnetization profiles especially the surface magnetization
$m_1$ as a function of $h_1$ can be extracted. It is instructive
to compare the results for the 3-$d$ Ising model with those for
the two-dimensional case obtained in Ref.\,\cite{twod}. This
is done in Fig.\,2. The crosses represent the data obtained
from a three-dimensional system with $N=256$, the circles stem
from the two-dimensional Ising model with lattice size
512$\times$2048.   
In both cases the statistical errors are
smaller than the symbol size.
\\[2mm]  
\begin{figure}[h]
\def\epsfsize#1#2{0.55#1}
\hspace*{2cm}\epsfbox{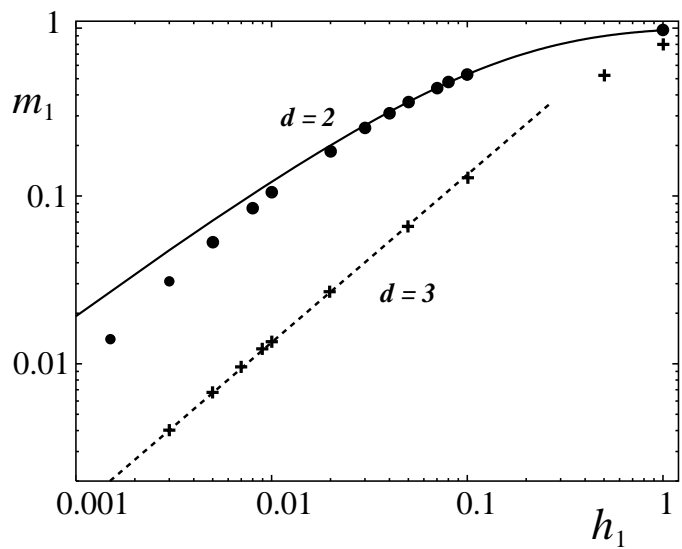}
\caption{Surface magnetization $m_1$ as a function of
$h_1$ for the Ising model in $d=2$ (full circles) and $d=3$ (crosses)
in double-logarithmic representation. The data for $d=2$ stem from
a 512$\times$2048 lattice, those for $d=3$ from a $256^2\times 512$ system.
The solid line shows the exact result for the $d=2$ semi-infinite
system. The dashed line is a fit
to the our MC data.}

\end{figure}

The situation in three dimensions
is obviously quite simple. Up to $h_1\simeq 0.1$ the response
of $m_1$ on $h_1$ is just linear. 
This is the regime where the
scaling analysis of Sec.\,\ref{scalan} applies, in particular
the basic assumption that $m(z)\sim h_1$ as $z$ goes to zero.
The dashed line is a
linear fit to the data for $h_1\le 0.1$.
For larger values of $h_1$ the surface magnetization saturates, such
that for $h_1\to \infty$ the curve has to approach unity

The dependence of $m_1$ on the surface field in two dimensions
is more complicated. As discussed in detail in Ref.\,\cite{twod}
and investigated in many exact calculations  \cite{fishau}, there
occurs a logarithmic factor in the functional dependence of $m_1$ on $h_1$. 
The solid line shows the exact result of the semi-infinite system, which,
due to the logarithm, never goes through a regime of linear behavior. The MC data
(see Ref.\,\cite{twod}) deviate from the exact curve for
$h_1\lesssim 0.02$ and for small $h_1$ indeed show a linear
dependence. This observation, linear response for small $h_1$ and an approach to the
true semi-infinite behavior for larger $h_1$, is a finite-size effect
consistent with exact results  \cite{fishau}.
  
\begin{figure}[b]
\def\epsfsize#1#2{0.5#1}
\hspace*{2cm}\epsfbox{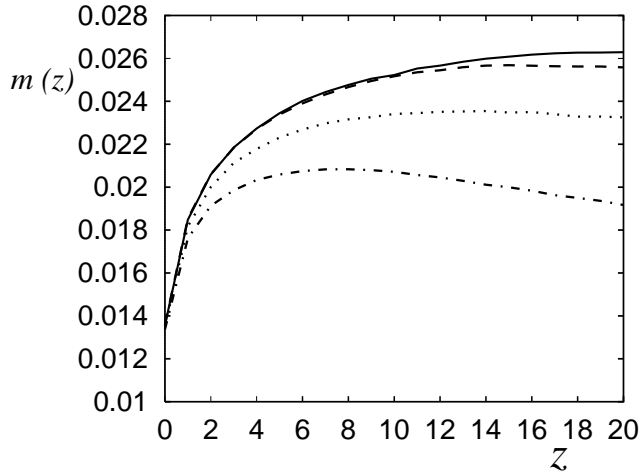}
\caption{The local magnetization at and in the immediate vicinity
of the surface for $h_1=0.005$ and $N=64$ (dashed-dotted),
128 (dotted), 196 (dashed), and 256 (solid line).}
\end{figure}

The next point are finite-size effects in the vicinity of the surface, especially
concerning the dependence of $m_1$ on $N$. As discussed in Sec.\,\ref{fisi},
we expect that $m_1$ and the profile up to a certain distance $\sim N$
should not depend on $N$. 
In Fig.\, 3 we plotted the data for
the local magnetization for four different system sizes ranging
from $N=64$ to 256 up to $z=20$.
In all cases the surface field was  $h_1=0.01$.
Quite obviously, $m_1$ itself does not vary with $N$
in the given range of sizes, confirming the finite-size scaling
analysis in Sec.\,\ref{fisi} as well as  the assumptions underlying
the scaling analysis in Sec.\,\ref{scalan}.  
From the given value of $m_1$
all profiles increase for $z$ increasing away from the surface. For the first few
layers the curves lie on top of each other, but in the smaller systems 
the slopes become smaller compared to larger $N$ at relatively small
distances already. For the
system with $64^2\times 128$ spins the regime with growing magnetization
extends to $z\simeq 7$ only.

\begin{figure}[t]
\def\epsfsize#1#2{0.5#1}
\hspace*{2cm}\epsfbox{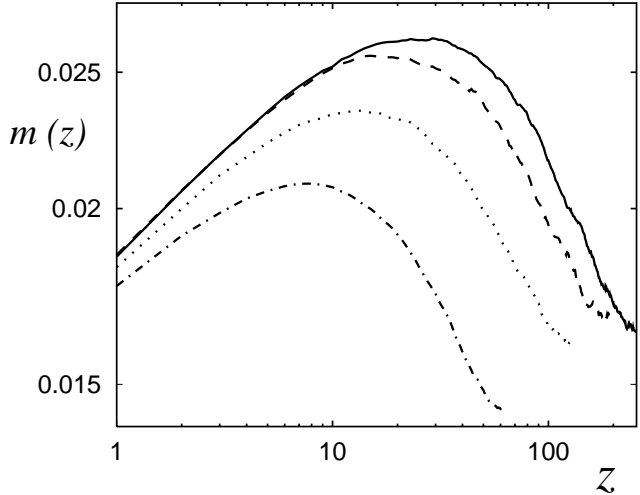}
\caption{Order-parameter profiles for the same parameters as in Fig.\,3
in double-logarithmic representation.}
\end{figure}

The variation of the same curves on larger scales is 
displayed in Fig.\,4 in double-logarithmic form. 
Going to larger system size the distance of the maximum $z_{max}$
grows roughly as $\sim N$. For $N=256$ we have $z_{max}\simeq 30$.
Recalling the results of the finite-size scaling analysis of
Sec.\,\ref{fisi}, we conclude that with these parameters the model
is in the regime where $L<l^{ord}$, and going to a smaller $h_1$ that
would increase $l^{ord}$ would not help to extend the region
of growing magnetization.
For $z\to 0$ the form of the profiles is consistent
with a power law. However, in the small systems the finite
size effects cause the profiles to crossover to the exponential
decay (compare Sec.\,\ref{fisi}) at a rather small distance.
Even in the largest system $N=256$ that could be studied by our
present means
with reasonable effort,
the near-surface power law does not extend beyond
$z\simeq 20$. The problem is that we indeed have $L\sim N$, but
apparently with a rather small constant of proportionality.

Nevertheless a rough value for the
short-distance exponent can be extracted from
the profile for $N=256$. The result is $\kappa=0.16(2)$. This
is somewhat lower than our expectation. The deviation from the expected
value $0.21$ is very likely due to the finite-size effects. We can not claim to
see the power law (\ref{power}) over a really macroscopic range before
the crossover to the finite-size (exponential) behavior sets in. So
the determination of a more reliable value of $\kappa$ from the
short-distance behavior remains as a task for larger-scale simulations.
 
\begin{figure}[b]
\def\epsfsize#1#2{0.5#1}
\hspace*{2cm}\epsfbox{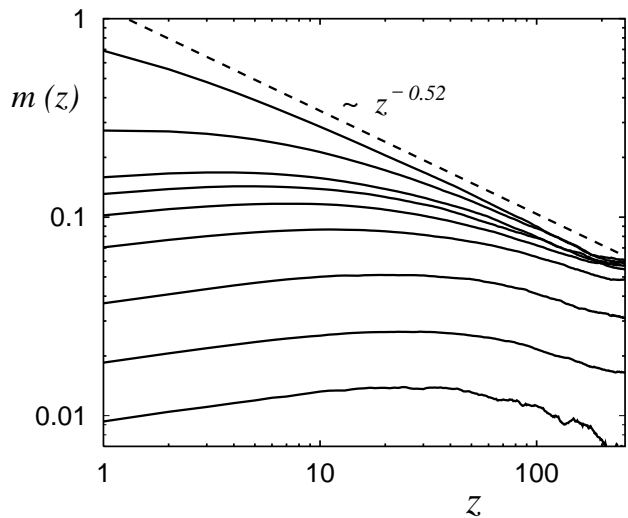}
\caption{Order-parameter profiles for $N=256$ and $h_1=0.005$,
0.01, 0.02, 0.04, 0.08, 0.1, 0.2, and 5.0 in double-logarithmic representation.
The pure power law $\sim z^{-0.52}$ (dashed line) is shown for comparison.}
\end{figure}
 
\begin{figure}[b]
\def\epsfsize#1#2{0.5#1}
\hspace*{2cm}\epsfbox{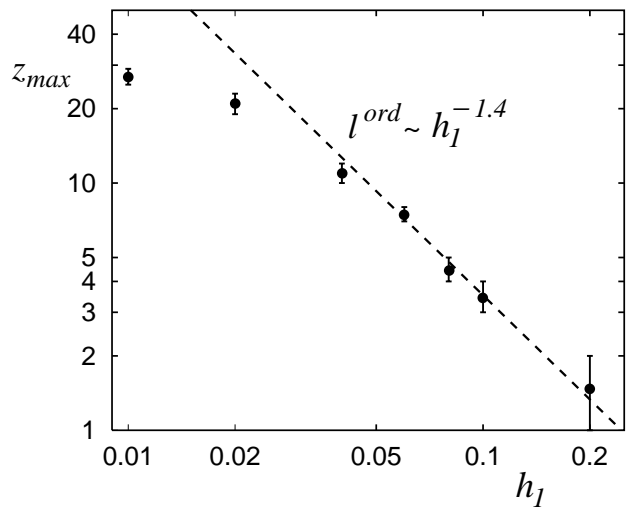}
\caption{$z_{max}$
in dependence of $h_1$ in double-logarithmic representation as
obtained from the profiles of Fig.\,5.  
The error bars were estimated. The dashed line depicts the derived
power-law dependence of the length scale $l^{ord}$ on $h_1$.}
\end{figure}

Magnetization profiles for $N=256$ and $h_1$ varying
in a wide range between 0.005 and 5 are plotted in Fig.\,5.
The dashed line represents the pure power law $\sim z^{-0.52}$
characteristic for the extraordinary or normal transition.
For small $h_1$, up to $h_1\simeq 0.1$, the curves show the near-surface growth
consistent with (\ref{power}).
For $h_1$ up to about 0.02, the location of the
maximum (here $z_{max}\simeq 30$) is determined by the finite-size scale $L$.
Setting   $h_1$ to larger values, the maximum moves closer to the surface. This is the regime
where $l^{ord}$ is smaller than $L$ and the location of the maximum is
governed by $l^{ord}$. 
In the case of  $h_1\gtrsim 0.2$, the profiles decay monotonously. Setting
$h_1=5.0$, the magnetization at the surface is
very close to one, and the decay 
for $10\lesssim z\lesssim 100$ is consistent with the power law $\sim z^{-x_{\phi}}$
The value of the exponent $x_{\phi}=\beta/\nu$ obtained from
this curve is $0.51(1)$, which is in excellent agreement with the literature value
0.517  \cite{ferlan}. The up-bending of the profiles for $z\gtrsim 100$
is due to the second surface.
   
Eventually, Fig.\,6 shows the values of $z_{max}$ as a function of $h_1$ determined
from the profiles of Fig.\,5.
From these data, the qualitative picture from above can be made more quantitative.
Especially with the result Eq.\,(\ref{length}) we can in principle determine directly
the scaling dimension $y_1^{ord}$. 
A power law fitted to the data for $0.02\le h_1\le 0.1$ yields $1/y_1^{ord}=1.4 (1)$
(where the error was estimated), which, in turn, yields 0.71(4) for
the scaling dimension $y_1^{ord}$. As mentioned above the
literature value is 0.73. However,
as in the case of the short-distance exponent, we also here have to admit
that we are not really in a regime where we can call $l^{ord}$ large
compared with the lattice spacing, and the good agreement
with what we expected from the scaling analysis is actually
surprising.

\section{Summary and Concluding Remarks}\label{summary}
We studied the near-surface behavior of the
order parameter in the three-dimensional Ising system
under the influence of a surface magnetic field $H_1$.
The anomalous behavior
found in \cite{czeri} by employing scaling arguments and
perturbative methods
was confirmed in the present work by Monte Carlo simulations.
Especially, the short-distance power law (\ref{power}) was
corroborated.

However, in our Monte Carlo study especially the region with small $H_1$ is
severely affected by finite size effects. Even in our largest system
($256^2\times 512$) the increase of $m(z)$ does not extend
beyond $z\simeq 30$. In order to obtain reliable results
for the exponent $\kappa$
determined with the help of the short-distance power law
(\ref{power}) and profiles that can be used for the quantitative
comparison with experimental data, one has to go to systems
beyond the size that we are able to treat by our present means.

Concerning experiments on surface critical phenomena, our
results should be of interest especially
in those cases where a small $H_1$ occurs,
at a surface that disfavors the order.
An example where this was obviously realized is the system studied by
Desai et al.  \cite{franck}.  In their experiment a binary fluid was studied in a container
whose walls as a function of time 
change their preference from one component to the other, the
time scale of this change being of the order of days. In other words,
the surface field $H_1$ changes sign as time goes by, and during a certain period $H_1$
is small.  First substantial
steps towards a theoretical explanation of this
and other similar experiments on binary mixtures
were made already by Ciach et al.  \cite{alina}.
A complete theoretical analysis would require a careful derivation
of experimentally observably quantities like reflectivity and ellipticity
for light scattering experiments on the basis of our results for the
order-parameter profile.

Another experiment,
discussed already in some detail in \cite{czeri}, is the
one by Mail\"ander et al.  \cite{mail} on Fe$_3$\,Al. This system undergoes
(among other transitions) a second-order phase transition
between a phase with $DO_3$ structure
and one with $B2$ structure. The near-surface regime
was studied by scattering of evanescent x-rays. The exponents observed were
consistent with the expectation for the ordinary transition, but Bragg peaks
revealed the existence of long-range order near the surface reminiscent to the
normal transition.
In order to explain the experimental results of  \cite{mail} on the basis
of our findings we have to assume that there exits an effective $H_1$
in this system. Then, if the associated length scale $l^{ord}$
is larger than both the scattering depth and the bulk correlation length,
the structure function is governed by the anomalous
dimension of the ordinary transition. On the other hand, the
steep increase of the order parameter should provide the explanation
for the observed long-range order near the surface. \\[5mm] 
{\small {\it Acknowledgements}: We thank A. Ciach for helpful comments
and discussions. We are especially indebted to our system administrator
R. Oberhage for his professional support concerning computer questions.
This work was supported in part by the Deutsche Forschungsgemeinschaft
through Sonderforschungsbereich 237.}

\end{document}